\documentclass[reprint,aps,pra,superscriptaddress,amsmath,amssymb]{revtex4-1}
\usepackage{graphicx}
\usepackage[colorlinks=true,allcolors=blue]{hyperref}

\usepackage{physics}
\usepackage{mathtools}
\usepackage{siunitx}
\usepackage{xspace}
\usepackage[dvipsnames]{xcolor}

\newcommand{\vect}[1]{\vb*{#1}}

\newcommand{\dr}{\dd{\vect{r}}}

\newcommand{\imag}{\mathrm{i}}
\newcommand{\e}{\mathrm{e}}
\newcommand{\hc}{\text{h.c.}}

\newcommand{\R}{\mathbb{R}}

\newcommand{\vF}{v_F}
\newcommand{\kB}{k_B}
\newcommand{\Tc}{T_c}
\newcommand{\kc}{k_c}
\newcommand{\epsilonc}{\epsilon_c}
\newcommand{\lambdac}{\lambda_c}

\newcommand{\ie}{\emph{i.e.}\xspace}

\newcommand{\moire}{moir\'e\xspace}

\begin{document}

\title{Mean-field theory for superconductivity in twisted bilayer graphene}

\author{Teemu J. Peltonen}
\author{Risto Ojaj\"arvi}
\author{Tero T.~Heikkil\"a}

\affiliation{Department of Physics and Nanoscience Center, University of Jyvaskyla, P.O. Box 35 (YFL), FI-40014 University of Jyvaskyla, Finland}

\date{\today}

\begin{abstract}
Recent experiments show how a bilayer graphene twisted around a certain magic angle becomes superconducting as it is doped into a region with approximate flat bands. We investigate the mean-field $s$-wave superconducting state in such a system and show how the state evolves as the twist angle is tuned, and as a function of the doping level. We argue that part of the experimental findings could well be understood to result from an attractive electron--electron interaction mediated by electron--phonon coupling, but the flat-band nature of the excitation spectrum makes also superconductivity quite unusual. For example, as the flat-band states are highly localized around certain spots in the structure, also the superconducting order parameter becomes strongly inhomogeneous.
\end{abstract}

\maketitle

\section{Introduction}

Experiments on strongly doped graphene \cite{ludbrook2015evidence,tiwari2017superconductivity,Chapman2016,ichinokura2016superconducting} have shown that with proper preparations, graphene can be driven to the superconducting state. Such experiments indicate that the lack of superconductivity in undoped graphene is not necessarily due to a lack of an (effective) attractive electron--electron interaction with strength $\lambda$ that would drive graphene superconducting, but rather the small density of states (DOS) close to the Dirac point. Technically, in contrast to the Cooper instability for metals taking place with arbitrarily small $\lambda$, superconductivity in an electron system with a massless Dirac dispersion $\epsilon_p^2 = \vF^2 p^2$ and an energy cutoff $\epsilonc$ has a quantum critical point $\lambdac=\pi \hbar^2 \vF^2/(2\epsilonc)$ \cite{Kopnin2008} such that for $\lambda < \lambdac$, mean-field superconductivity does not show up at any temperature. From this perspective, doping to a potential $\mu$ leads to an increased DOS, and thereby to a non-vanishing critical temperature $\Tc \approx |\mu| \exp[- (\lambdac/\lambda-1)\epsilonc/|\mu|-1]$. An alternative approach would be to change the spectrum and increase the density of states close to the Dirac point. The extreme limit would be an approximately flat band of size $\Omega_\text{FB}$, where the group velocity tends to zero. In such systems the critical temperature is a linear function of the coupling strength, $\Tc = \lambda \Omega_\text{FB}/\pi^2$ \cite{Heikkila2011,supplement}, and quite high $\Tc$ can be expected even without extra doping \cite{khodel1990superfluidity, kopnin2011high, tang2014strain, kauppila2016flat, heikkila2016flat, lothman2017universal}.

\begin{figure}
    \includegraphics[width=\columnwidth]{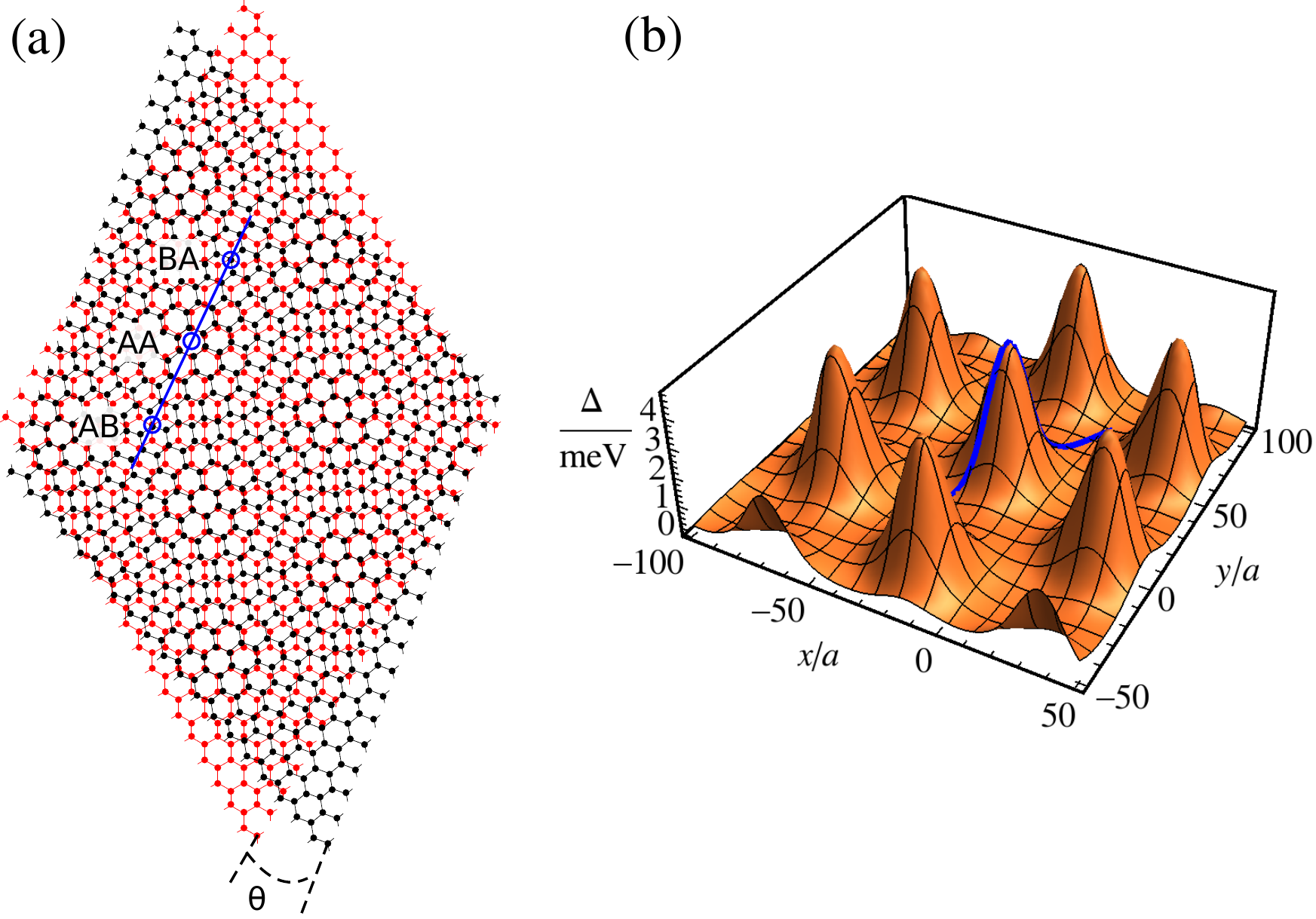}
    \caption{(a) Twisted bilayer graphene and its \moire superlattice. The upper layer is rotated by an angle $\theta$ relative to the lower layer. (b) Position dependence of the self-consistent $\Delta$, shown here at $T=0$ for the magic angle $\theta=\SI{0.96}{\degree}$ and $\lambda=\SI{5}{\eV}a^2$. In both figures also a line passing through high-symmetry points with AB, AA, and BA stacking is shown.}
    \label{fig:TBG+Deltavsr}
\end{figure}

Recent observations \cite{cao2018unconventional} of superconductivity in twisted bilayer graphene [TBG, see Fig.~\ref{fig:TBG+Deltavsr}(a)] take place in systems where theoretical studies have predicted the occurrence of asymptotically flat bands \cite{LopesdosSantos2007, SuarezMorell2010, Mele2010, mele2011band, bistritzer2011moire, LopesdosSantos2012, geim2013van, weckbecker2016low, fang2016electronic, ramires2018electrically, nam2017lattice}. There have been many suggestions of an unconventional superconducting state both for regular graphene \cite{uchoa2007,nandkishore2012chiral} and for TBG \cite{po2018origin, xu2018topological, ramires2018electrically, liu2018chiral, ray2018wannier, huang2018antiferromagnetically, dodaro2018phases, baskaran2018theory, roy2018unconventional, Guo2018}, typically directly related with the Coulomb interaction, and in some cases related with non-local interactions. Here we study the mean field theory of superconductivity in such systems, starting instead from the hypothesis that the observations could be explained with the conventional electron--phonon mechanism from the flat-band perspective \cite{Volovik2018}. This hypothesis is justified on the grounds that the relative strength and the screening of attractive and repulsive interactions are uncertain. Furthermore, phonon-mediated attraction is considered a viable mechanism for the observed superconductivity on doped graphene \cite{ludbrook2015evidence, tiwari2017superconductivity, Chapman2016, ichinokura2016superconducting, profeta2012}.

In particular, we use the model of Refs.~\onlinecite{LopesdosSantos2007,LopesdosSantos2012} for the spectrum of the twisted bilayer, add an on-site (leading to $s$-wave superconductivity) attractive interaction of strength $\lambda$, and evaluate the mean-field order parameter profile. We find that the order parameter, and along with it the mean-field critical temperature, have a similar non-monotonous behavior with respect to the twist angle as in the experiments. We also predict the behavior of the density of states in the superconducting state, resulting from the peculiarities of the flat-band eigenstates and from the position dependence of the superconducting order parameter [Fig.~\ref{fig:TBG+Deltavsr}(b)]. Even if our pairing interaction is quite simple, the resulting energy dependent density of states is quite unusual. In addition, we show how doping away from the flat band eventually destroys superconductivity.

\section{Normal state}

We describe the normal state of TBG with the model of Refs.~\onlinecite{LopesdosSantos2007,LopesdosSantos2012}. With this model, we can describe the twist angles $\theta$ at which the lattices \(L\) and \(L^\theta\) of the two graphene layers are commensurate, so that the system as a whole is periodic in the \moire superlattice \(SL\). Here we study only the simple commensurate structures, characterized by a single rotation parameter $m\in\mathbb{N}$, for which the rotation angle is given by
\begin{equation}
	\cos(\theta) = \frac{3m^2+3m+1/2}{3m^2+3m+1}.
\end{equation}
According to Ref. \onlinecite{LopesdosSantos2012} these structures approximate arbitrary commensurate structures. The primitive vectors of the superlattice $SL$ are given by
	$\vect{t}_1 = m\vect{a}_1 + (m+1)\vect{a}_2$, $\vect{t}_2 = -(m+1)\vect{a}_1 + (2m+1)\vect{a}_2$
and the primitive vectors of the reciprocal superlattice $SL^*$ are
	$\vect{G}_1 = \frac{4\pi}{3||\vect{t}_1||^2} ((3m+1)\vect{a}_1 + \vect{a}_2)$, 
    $\vect{G}_2 = \frac{4\pi}{3||\vect{t}_1||^2} (-(3m+2)\vect{a}_1 + (3m+1)\vect{a}_2)$,
where the lattice constant of the superlattice is $||\vect{t}_1|| = \sqrt{3m^2+3m+1}\,a$ and the graphene lattice primitive vectors are $\vect{a}_1=(1,\sqrt{3})a/2$ and $\vect{a}_2=(-1,\sqrt{3})a/2$ with $a$ the lattice constant \cite{LopesdosSantos2007}. In the following, we assume that \(\vect G \in SL^*\) belongs to the reciprocal superlattice, \(\vect k \in \R^2/SL^*\) to the first Brillouin zone of the superlattice, and also that the corresponding sums and integrals are restricted to these sets.

In the normal state, TBG is described by a low energy effective Hamiltonian \cite{LopesdosSantos2007}
\begin{align}
\label{eq:hamiltonian_normal}
&\mathcal{H}_{\rho\vect k}(\vect G,\vect G') = \\
&\mqty(
	[\hbar\vF \vect \sigma^\rho \cdot \left(\vect k + \vect G + \rho\Delta\vect{K}/2\right)-\mu]\delta_{\vect G, \vect G'}\mkern-110mu & \mkern+120mu t_\perp^\rho(\vect G - \vect G') \\\mkern-70mu
	t_\perp^\rho(\vect G' - \vect G)^\dag & \mkern-70mu  [\hbar\vF \vect \sigma^\rho_\theta \cdot \left(\vect k + \vect G - \rho\Delta\vect{K}/2\right)-\mu]
\delta_{\vect G, \vect G'}
),\notag%
\end{align}
where the matrix structure corresponds to the layer structure and \(\rho\in\{+,-\}\) is the valley index with $+$ corresponding to $\vect{K}$ and $-$ to $\vect{K}'=-\vect{K}$. Furthermore, each entry is a \(2\times 2\) matrix due to the sublattice structure in graphene. The diagonal terms in Eq.~\eqref{eq:hamiltonian_normal} describe the Dirac dispersion in the two layers and are diagonal also in \(\vect G\). Here, \(\vect{\sigma}^\rho=(\rho\sigma_x,\sigma_y)\). For the second layer we include the rotation \(\theta\) so that  \(\vect{\sigma}^\rho_\theta = \e^{+\imag\theta \sigma_z/2}\vect{\sigma}^\rho \e^{-\imag\theta \sigma_z/2}\). \(\Delta\vect{K}=\vect{K}^\theta - \vect{K}\) is the relative shift of the Dirac cones between the layers. The coordinates correspond to those of layer 1 as measured from the \(\vect K\)-point, but shifted with a vector \(+\Delta\vect{K}/2\) for layer 1 and \(-\Delta\vect{K}/2\) for layer 2. With this choice, the relative momentum \(\vect k\) on both layers corresponds to the same absolute momentum. Furthermore, \(\mu\) is the chemical potential describing the effect of doping, here taken to be equal in both layers.

\begin{figure}
    \includegraphics[width=\columnwidth]{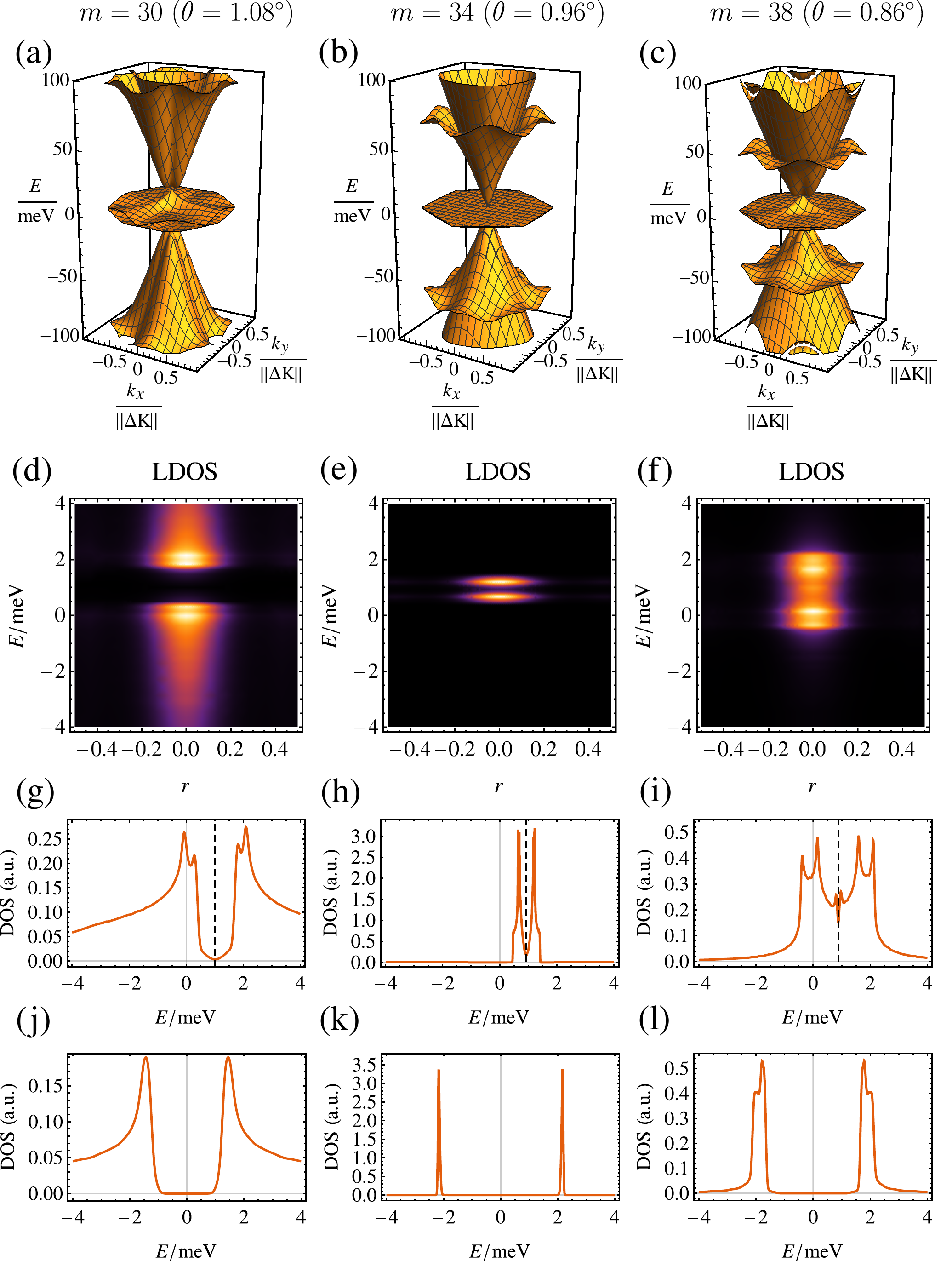}
    \caption{(a--c) Normal-state dispersion, (d--f) local and (g--i) total density of states for three different angles near the magic angle $\theta=\SI{0.96}{\degree}$ in the normal state. The bottom row (j--l) shows the corresponding total density of states in the superconducting state, in the case $T=0$ and $\lambda=\SI{5}{\eV}a^2$ and when doped to the point $\mu_0$ marked as a dashed line in (g--i).} 
    \label{fig:normalstate}
\end{figure}

The off-diagonal terms in the Hamiltonian describe the coupling between the two layers. The matrix element at valley $\rho$ between a state in sublattice \(\alpha\) in layer 1 and a state in sublattice \(\beta\) in layer 2 is
\begin{equation}
t_\perp^{\rho,\alpha\beta}(\vect G) = \frac{1}{N} \sum_{\vect{r}} \e^{-\imag\vect{G}\cdot(\vect{r}+\delta_{\alpha B}\vect{\delta}_1)} \e^{\imag\rho\vect{K}^\theta\cdot\vect{\delta}^{\alpha\beta}(\vect{r})} t_\perp(\vect{\delta}^{\alpha\beta}(\vect{r})),
\end{equation}
where $\vect{\delta}^{\alpha\beta}(\vect{r})$ is the horizontal displacement vector between the site at $\vect{r}$, sublattice $\alpha$ in layer 1 and the nearest neighbor at sublattice $\beta$ in layer 2. $\vect{\delta}_1$ denotes one of the nearest neighbor vectors connecting the graphene A and B sublattices. The sum is over the graphene A sublattice sites in the superlattice unit cell, and $N$ denotes the number of these sites. For the interlayer hopping energy $t_\perp(\vect{\delta})$ we use the same Slater--Koster parameterization as in Ref.~\onlinecite{LopesdosSantos2007}. Furthermore, we approximate the interlayer coupling by only including the matrix elements with $\vect G\in\{0,-\vect{G}_1,-\vect{G}_1-\vect{G}_2\}$ (valley $\vect{K}$) or $\vect G\in\{0,\vect{G}_1,\vect{G}_1+\vect{G}_2\}$ (valley $\vect{K}'$), since they are an order of magnitude larger than the rest.

For $\theta \approx \SI{1}{\degree}$, the electronic dispersion becomes almost flat \cite{bistritzer2011moire} and the group velocity $\dd\epsilon_p/\dd p$ tends towards zero. In Fig.~\ref{fig:normalstate} we plot the resulting normal-state dispersion (a--c) and the (local and total) density of states (d--i) close to this ``magic'' angle. The exact value of this magic angle depends on the details of the hopping model. In our case it is around $\SI{0.96}{\degree}$, \ie, somewhat lower than what was found in Ref.~\onlinecite{bistritzer2011moire}. However, the qualitative behavior of the local density of states is rather similar to the previous models. In particular, there are two closely spaced DOS peaks signifying the flattening of the bands. The local density of states is plotted along the line shown in Fig.~\ref{fig:TBG+Deltavsr}, including three high-symmetry points with AB, AA, and BA stacking. These correspond to $r=-1/3$, $0$, and $1/3$, respectively.

\section{Superconducting state}

We assume that there is a local attractive interaction $\lambda_{\sigma_1\sigma_2}(\vect{r}_1,\vect{r}_2) = \delta_{\bar{\sigma}_1\sigma_2}\delta(\vect{r}_1-\vect{r}_2)\lambda$ with strength \(\lambda\), which results \cite{supplement} in an order parameter $\Delta_{\alpha i}(\vect{r})$ depending only on the center-of-mass coordinate $\vect{r}$ (and sublattice $\alpha$ and layer $i$). On the other hand the classification of the order parameter symmetries to $s$, $d$, $f$, etc. is based only on the relative coordinate $\vect{r}_1-\vect{r}_2$, which in our model is always zero. Thus the symmetry is purely $s$-wave.

\begin{figure}
   \includegraphics[width=\columnwidth]{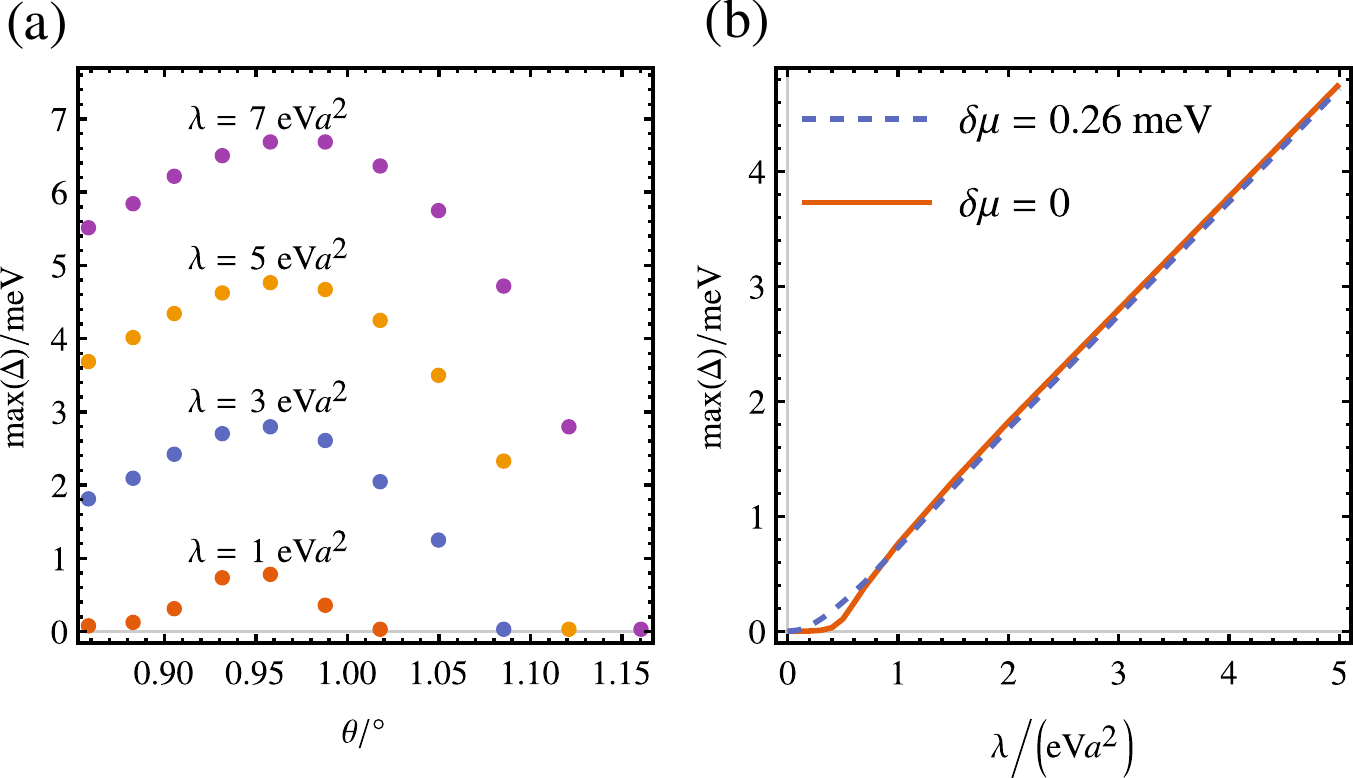}
    \caption{Maximum of the position-dependent superconducting order parameter $\Delta(\vect{r})$ at $T=0$ as a function of (a) the rotation angle and (b) the coupling strength for $\theta=\SI{0.96}{\degree}$. In (b) we also show how doping to the DOS peak affects the small-$\lambda$ behavior.}
    \label{fig:Deltavstheta+Deltavslambda}
\end{figure}

\begin{figure}
	\includegraphics[width=0.6\columnwidth]{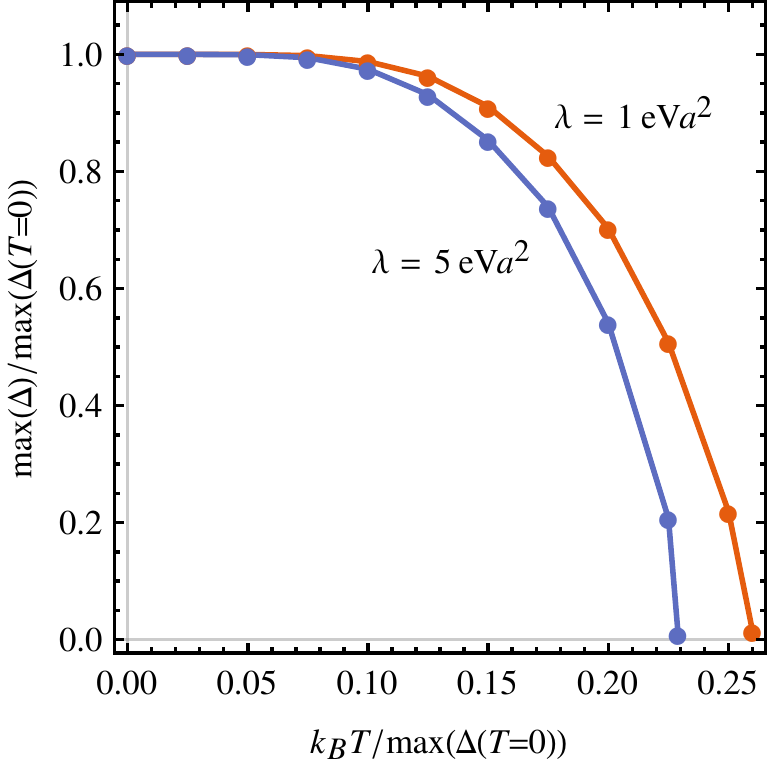}
    \caption{$\max(\Delta)$ as a function of temperature in the case $\theta=\SI{0.96}{\degree}$ for two values of $\lambda$, showing the approximate linear relation $\kB\Tc \approx {0.25}\,{\max(\Delta(T=0))}$ for the critical temperature. The dots are the calculated values and the lines are a guide to the eye.}
    \label{fig:DeltavsT}
\end{figure}

\begin{figure}
    \includegraphics[width=\columnwidth]{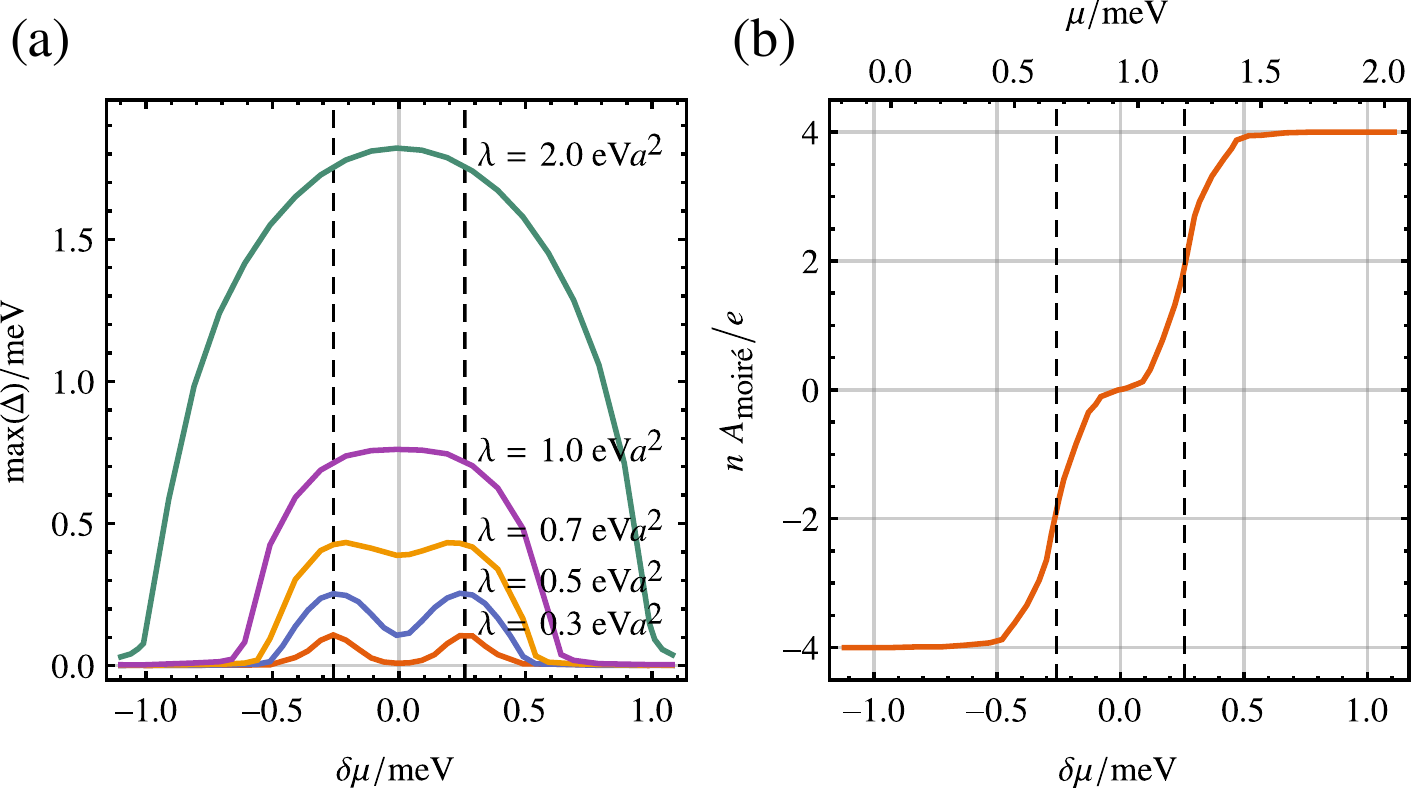}
    \caption{Effects of electrostatic doping $\mu=\mu_0+\delta\mu$ for $\theta=\SI{0.96}{\degree}$. (a) $\max(\Delta)$ vs. chemical potential for various values of $\lambda$ at $T=0$. (b) Charge density in the normal state at $T=0$ as a function of chemical potential. The units of the charge density $n$ are $e/A_\text{\moire}$, where $e$ is the electron charge and $A_\text{\moire}$ is the area of the \moire unit cell. In both figures the vertical dashed lines mark the location of the DOS peaks at $\delta\mu \approx \pm\SI{0.26}{\meV}$.}
    \label{fig:chargedensity+Deltavsdmu}
\end{figure}

We do not consider the specific nature of the pairing interaction and for the purposes of this paper it can be mediated by phonons or other bosonic modes. This model disregards the retardation effects due to such a mechanism, but is a valid approximation to the more general Eliashberg approach for weak coupling \cite{eliashberg1960interactions,Ojajarvi2018}. That theory also shows that direct Coulomb interaction, typically modeled via the Hubbard model, is less effective in reducing \(\Delta\) than what could be naively expected, and should be included in the low-energy self-consistency equation as a Coulomb pseudopotential \(u^* = u/(1+u\alpha)\) \cite{supplement,morel1962calculation,Ojajarvi2018}, where \(u=Ua^2\), \(U\) is the Hubbard interaction constant, and \(\alpha\) is a constant measuring the amount of renormalization due to the high energy bands above the electron--phonon cutoff frequency \(\omega_D\). For TBG we find from a simplified model \(\alpha\approx \SI{0.2}{eV^{-1}}a^{-2}\) \cite{supplement}. Thus, a combination of electron--phonon and Coulomb interactions leads to an effective interaction strength $\lambda_\text{eff} = \lambda - u^*$. As long as $\lambda_\text{eff}>0$, there is a possibility for a superconducting state even if \(u>\lambda\). For example, for $U=\SI{5}{eV}$, $u^*=\SI{2.5}{eV}a^2$ is in the same regime as the value of $\lambda_\text{eff}$ in Figs. \ref{fig:Deltavstheta+Deltavslambda}--\ref{fig:chargedensity+Deltavsdmu}. Note that in what follows, we refer to this $\lambda_\text{eff}$ simply as $\lambda$.

Within a mean-field theory in the Cooper channel we find a self-consistency equation for a
local superconducting order parameter \cite{supplement}. Assuming that this order
parameter shares the periodicity of the \moire superlattice, we find
the self-consistency equation
\begin{align}
	\Delta_{\alpha i}(\vect{G}) = &\lambda \sum_{\rho,b} \sum_{\vect{G}'} \int\!\frac{\dd{\vect{k}}}{(2\pi)^2} \tanh(\frac{E_{\rho b\vect{k}}}{2\kB T}) \notag\\
    &\times u_{\rho b\vect{k},\alpha i}(\vect{G}') v_{\rho b\vect{k},\alpha i}^*(\vect{G}'-\vect{G}),
\label{eq:self-consistency_equation_kspace}
\end{align}
where the band sum $b$ is calculated over the positive energy bands, \(\alpha\in\{A,B\}\) is the sublattice index, \(i\in\{1,2\}\) is the layer index, and \(u_{\rho b\vect{k}}\) and \(v_{\rho b\vect{k}}\) are the eigenvectors of the Bogoliubov--de Gennes equation
\begin{align}
\sum_{\vect{G}'}
    \begin{pmatrix}
		\mathcal{H}_{\rho\vect k}(\vect G,\vect G') & {\Delta}(\vect{G}-\vect{G}') \\
        {\Delta}^*(\vect{G}'-\vect{G}) & -\mathcal{H}_{\rho\vect k}(\vect G,\vect G')
	\end{pmatrix}
    \begin{pmatrix}
    		u_{\rho b\vect{k}}(\vect G') \\ v_{\rho b\vect{k}}(\vect G')
    \end{pmatrix}\notag\\
    =
	E_{\rho b\vect{k}}
	\begin{pmatrix}
		u_{\rho b\vect{k}}(\vect G) \\ v_{\rho b\vect{k}}(\vect G)
	\end{pmatrix}.\qquad
\label{eq:bdg_kspace}
\end{align}

We solve this self-consistent order parameter with a few values of the interaction constant $\lambda$ and for a few different twist angles $\theta$ close to the magic angle. We include in the sum the energy levels closest to zero energy. We have checked that the results are not sensitive to the value of the energy cutoff, which we implement as a cutoff in the $b$ and $\vect{G}$~sums. For comparison between different angles, we measure the chemical potential from $\mu_0$, corresponding to the charge neutrality and marked in Figs.~\ref{fig:normalstate}(g--i) with a dashed line, by writing $\mu=\mu_0+\delta\mu$. The chemical potential shift $\mu_0$ is caused by the interlayer coupling. Unless otherwise stated, all the results concern the behavior at $\delta\mu=0$. The resulting total density of states is plotted in Fig.~\ref{fig:normalstate}(j--k), to allow for a comparison to the normal state. The corresponding local density of states (not shown) has the same localized structure as in the normal state, but the energy dependence is modified similarly as the total DOS. The effect of finite temperature on the superconducting (L)DOS happens solely via $\Delta(T)$, which is calculated below.

The maximum of the position dependent $\Delta$, which according to numerics is equal in both layers and sublattices, is plotted in Fig.~\ref{fig:Deltavstheta+Deltavslambda}(a) for different values of the twist angle and for four different coupling strengths. The precise angle for the maximum depends a bit on the chosen coupling strength. Moreover, $\max(\Delta)$ is almost a linear function of $\lambda$ [see Fig.~\ref{fig:Deltavstheta+Deltavslambda}(b)], as appropriate for a flat-band superconductor \cite{Heikkila2011}. This linearity is even more pronounced when the system is doped to the DOS peak at $\delta\mu\approx\SI{0.26}{\meV}$. Far from the magic angle, the Fermi speed $\vF(\theta)$ increases so that the chosen $\lambda$ is below the critical value $\lambdac$. This is why $\Delta$ vanishes for angles $\theta \gtrsim \SI{1.1}{\degree}$. 

We can analyze the resulting magnitude of $\Delta$ based on a
flat-band result (assuming a position independent $\Delta$ and $E_{\rho b\vect{k}} \approx \Delta$ for an extreme flat band) according to which $\Delta=\lambda \Omega_\text{FB}/\pi^2$ \cite{supplement}, where
$\Omega_\text{FB} \approx \Omega_\text{\moire} = 8 \pi^2/(\sqrt{3}||\vect{t}_1||^2)$. This yields $\Delta = 1.3 \times 10^{-3} \lambda/a^2$ for
$m=34$ corresponding to the magic angle. For comparison a linear fit to the linear region in Fig.~\ref{fig:Deltavstheta+Deltavslambda}(b) gives $\max(\Delta) = \SI{-0.2}{\meV}+1.0\times 10^{-3}\lambda/a^2$. The magnitude hence agrees very well with this simple model. Note that the precise values of these parameters especially for small $\lambda$ depend on the exact value of doping as shown below.

In Fig.~\ref{fig:DeltavsT} we show the temperature dependence of $\Delta$ for $m=34$, from which we may infer the approximate linear relation $\kB\Tc \approx 0.25\max(\Delta(T=0))$ for the critical temperature. The prefactor is somewhat lower than for an extreme flat band with a constant $\Delta$, for which \cite{supplement} $\kB\Tc = \Delta/2$. The difference is most likely explained by the nonvanishing bandwidth and the position dependent $\Delta$ of our model. The maximum critical temperatures for the models calculated in Fig.~\ref{fig:Deltavstheta+Deltavslambda}(a) range from $\SI{3}{\kelvin}$ to about $\SI{20}{\kelvin}$. The lower end of these values, calculated with $\lambda = \SI{1}{\eV}a^2$, is thus quite close to that found in Ref.~\onlinecite{cao2018unconventional}.

We stress that the above result is the mean-field critical temperature; the observed resistance transition is most likely rather a Berezinskii--Kosterlitz--Thouless (BKT) transition \cite{berezinskii1972destruction,kosterlitz1973ordering}. Therefore, the mean-field $T_c$ gives an upper bound for the measured transition temperature. Furthermore, even the BKT transition temperature can be calculated from the mean-field superfluid weight \cite{Julku2018}. The mean-field results are also relevant in that the (L)DOS can be experimentally measured by tunneling experiments and this depends on the structure and magnitude of mean-field \(\Delta\) at temperatures below the BKT transition. Note that despite the flatness of the bands, the supercurrent can be non-vanishing in the case when the eigenstate Wannier functions overlap \cite{peotta2015superfluidity} as is the case for TBG.

Besides $\theta$-dependence, we can check how doping away from the center of the two DOS peaks affects the superconducting state. In Fig.~\ref{fig:chargedensity+Deltavsdmu}(a) we plot the order parameter $\max(\Delta(\delta\mu))$ for different values of the doping $\delta\mu$ as measured from the charge neutrality point. Close to the magic angle, for $\lambda \gtrsim \SI{1}{eV}a^2$ the energy scale of superconductivity exceeds that of the normal-state dispersion, and hence the only effect of the doping is to move away from the flat-band regime, suppressing superconductivity \cite{kopninbook2014}. For smaller values of $\lambda$, $\max(\Delta)$ is smaller than the bandwidth, and hence doping to the DOS peaks enhances superconductivity. Especially for $\lambda\lesssim\SI{0.3}{eV}a^2$ there are separate superconducting domes with doping levels close to the DOS peaks, which resembles the phase diagram in Ref.~\onlinecite{cao2018unconventional} for hole ($n<0$) doping, apart from the insulating state at $n\approx -2 e/A_\text{\moire}$. For electron doping ($n>0$), superconductivity is absent in the experiment, whereas our model is electron--hole symmetric. Since Ref.~\onlinecite{cao2018unconventional} uses charge density $n$ as a unit for the doping level while our theory is formulated in terms of the chemical potential $\mu$, for easier comparison we show the dependence between the charge density \cite{supplement} and chemical potential in Fig.~\ref{fig:chargedensity+Deltavsdmu}(b). From the figure we find that the DOS peaks correspond to approximately $\pm 2$ extra electrons per \moire unit cell.

\section{Conclusions}

Concluding, we find that a BCS-type mean field model with relatively weak attractive interaction constant possibly even due to electron--phonon coupling can explain the occurrence of superconductivity in twisted bilayer graphene. We also make a number of predictions concerning the mean-field superconducting state, in particular the density of states and doping dependence. Our results form hence a checkpoint for further studies, that use a simplified picture of the TBG flat-band states or consider mechanisms beyond the one in this paper. Our results could also have relevance in explaining the observations of superconductivity in twisted interfaces of graphite \cite{esquinazi2013graphite,Ballestar2013,stiller2018local}.

Our mean-field theory fails to explain the insulator state \cite{padhi2018wigner} found experimentally in TBG at $n \approx \pm2e/A_\text{\moire}$ as well as the lack of superconductivity for electron doping \cite{cao2018unconventional,cao2018correlated}. However, the latter of these cannot be seen as a drawback of our model as in another experiment \cite{Yankowitz2018} some samples were found to be superconducting also on the electron-doped side, and thus it clearly depends on the samples and on the experimental setup. Regarding the insulator phase, it is plausible that the mean-field theory fails when the doping level corresponds to an integer number of electrons per superlattice unit cell. The biggest discrepancy is however most likely caused by the possible dependence of $\lambda_\text{eff}$ on the charge density, because the effect of Coulomb interaction depends on charge screening. Within the flat-band model of Ref.~\onlinecite{Ojajarvi2018}, the case $\lambda_\text{eff}>0$ corresponds to a superconducting state, whereas for $\lambda_\text{eff}<0$ an insulating antiferromagnetic state is realized. Thus, by taking the chemical potential dependence of $\lambda_\text{eff}$ into account, it may be possible to describe both superconducting and insulating phases found in the experiment \cite{cao2018unconventional}. A detailed description would require generalizing Refs. \onlinecite{Ojajarvi2018} and \onlinecite{hwang2010plasmon} to the TBG case.

We point out that our simple BCS model disregards the strain effects in \moire bands, as well as the possible dependence of the interaction constant on the twist angle and doping level. Whereas such mechanisms may play a role in TBG, we believe that the simplest BCS-type mean field superconductivity should also be considered as a viable effective model of the observations. Nevertheless, even in this case superconductivity would be highly exceptional, for example because it can be so strongly controlled by electrostatic doping. 

\emph{Note added}. Soon after we submitted our work, Ref. \onlinecite{wu2018theory} addressed a similar BCS-type model as here, obtaining consistent results with this paper. In addition to local interactions leading to $s$-wave superconductivity, they considered also nonlocal interactions opening the possibility to $d$-wave superconductivity. They found out that without including Coulomb repulsion the $s$-wave channel is more stable, having a higher $\Tc$.

\begin{acknowledgments}
This project was supported by Academy of Finland Key Project funding, and the Center of Excellence program (Projects No. 305256 and 284594). We acknowledge grants of computer capacity from the Finnish Grid and Cloud Infrastructure (persistent identifier urn:nbn:fi:research-infras-2016072533).
\end{acknowledgments}

\bibliography{references}

\appendix

\section{Derivation of the self-consistency equation}
The Hamiltonian for a local attractive interaction of strength $\lambda>0$ is
\begin{equation}
	H_\text{int} = -\frac{\lambda}{2} \sum_{\mathclap{\sigma,\alpha,i\,}} \int\dr \psi_{\sigma,\alpha i}^\dagger(\vect{r}) \psi_{\bar\sigma,\alpha i}^\dagger(\vect{r})	\psi_{\bar\sigma,\alpha i}(\vect{r}) \psi_{\sigma,\alpha i}(\vect{r}),
	\label{eq:H_int}
\end{equation}
where $\psi_{\sigma,\alpha i}(\vect{r})$ is the annihilation operator for spin $\sigma$ at position $\vect{r}$, layer $i\in\{1,2\}$, and sublattice $\alpha\in\{A,B\}$. Doing the mean field approximation in the Cooper channel, assuming only intervalley coupling, and transforming to the valley operators by $\psi_{\sigma,\alpha i}(\vect{r}) = \sum_\rho \e^{\imag\rho\vect{K}\cdot\vect{r}} \psi_{\sigma\rho,\alpha i}(\vect{r})$ the interaction Hamiltonian becomes
\begin{align}
	H_\text{int} &= \frac{1}{2} \sum_{\sigma,\rho,\alpha,i} \int\dr \Delta_{\sigma,\alpha i}(\vect{r}) \psi_{\sigma\rho,\alpha i}^\dagger(\vect{r}) \psi_{\bar\sigma\bar\rho,\alpha i}^\dagger(\vect{r}) +\hc \notag\\
	&+\frac{1}{2\lambda} \sum_{\sigma,\alpha,i} \int\dr \abs{\Delta_{\sigma,\alpha i}(\vect{r})}^2,
    \label{eq:H_int_MF}
\end{align}
where the local superconducting order parameter is $\Delta_{\sigma,\alpha i}(\vect{r}) = -\lambda\sum_\rho \expval{\psi_{\bar\sigma\bar\rho,\alpha i}(\vect{r}) \psi_{\sigma\rho,\alpha i}(\vect{r})}$. Then by moving to the Nambu space and doing the Bogoliubov transformation we find that the self-consistency equation for the up-spin $\Delta_{\alpha i} \coloneqq \Delta_{\uparrow,\alpha i}$ becomes
\begin{align}
	\Delta_{\alpha i}(\vect{r}) = \lambda \sum_{\rho,b} \int\!&\frac{\dd{\vect{k}}}{(2\pi)^2} \tanh(\frac{E_{\rho b \vect k}}{2\kB T}) \notag\\
	&\times u_{\rho b\vect{k},\alpha i}(\vect{r}) v_{\rho b\vect{k},\alpha i}^*(\vect{r}),
    \label{eq:self-consistency_equation_realspace}
\end{align}
where \(u_{\rho b\vect{k},\alpha i}\) is the \((\alpha,i)\)-component of the spinor \(u_{\rho b\vect{k}}\) and the $b$ sum goes over the positive energy bands. The spinors \(u_{\rho b\vect{k}}\) and \(v_{\rho b\vect{k}}\) are determined by solving the Bogoliubov--de Gennes equation
\begin{equation}
    \begin{pmatrix}
    	\mathcal{H}_\rho(\vect{r}) & \Delta(\vect{r}) \\
        \Delta^*(\vect{r}) & -\mathcal{H}_\rho(\vect{r})
    \end{pmatrix}
    \begin{pmatrix}
    		u_{\rho b\vect{k}}(\vect{r}) \\ v_{\rho b\vect{k}}(\vect{r})
    \end{pmatrix}
=
	E_{\rho b\vect{k}}
	\begin{pmatrix}
		u_{\rho b\vect{k}}(\vect{r}) \\ v_{\rho b\vect{k}}(\vect{r})
	\end{pmatrix},
\label{eq:bdg_real_space}
\end{equation}
where $\Delta$ is a diagonal 4 by 4 matrix including the components $\Delta_{\alpha i}$.
Substituting the Bloch wave expansion
\begin{equation}
	\mqty(u_{\rho b\vect{k}}(\vect{r})\\ v_{\rho b\vect{k}}(\vect{r})) 
	= \e^{\imag\vect{k}\cdot\vect{r}} \sum_{\mathclap{\vect{G}'}} \e^{\imag\vect{G}'\cdot\vect{r}} 
	\mqty({u}_{\rho b\vect{k}}(\vect{G}')\\ {v}_{\rho b\vect{k}}(\vect{G}'))
    \label{eq:u_bloch_fourier_series}
\end{equation}
for the eigenstates into Eq.~\eqref{eq:bdg_real_space} and assuming \(\Delta(\vect r)\) to be periodic in the superlattice, we find the Fourier space Bogoliubov--de Gennes equation [Eq.~\eqref{eq:bdg_kspace} in the main text] and the Fourier space version of the self-consistency equation [Eq.~\eqref{eq:self-consistency_equation_kspace} in the main text]. 

\section{Charge density}

The non-coupled system of twisted bilayer graphene is charge neutral at the chemical potential \(\mu=0\). The charge density due to the electrons at that potential is
\begin{equation}
n_0 = \frac{2e}{V} \sum_{b\in B,\vect k} f_0(\epsilon_{0,b\vect k}) = \frac{2e}{V} \sum_{b\in\Omega,\vect k} f_0(\epsilon_{0,b\vect k}) + n_{\rm high}
 \label{eq:non-int_charge}
\end{equation}
with
\begin{equation*}
 n_{\rm high} = \frac{2e}{V} \sum_{b\in B\setminus\Omega,\vect k} f_0(\epsilon_{0,b\vect k}),
\end{equation*}
where \(e\) is the electron charge and the factor of 2 comes from the spin. We formulate the calculation so that the \(\vect k\)-sum goes over the superlattice Brillouin zone \(L_{BK}^*/SL^*\), \(B\) is the set of bands and \(\epsilon_{0,b\vect k}\) is the non-interacting dispersion. \(f_0\) is the Fermi-Dirac distribution function at zero temperature. In the second step we introduce a cutoff by dividing the sum over the bands into two terms; to a sum over a set of low-energy bands \(\Omega\) and to a sum over high-energy bands \(B\backslash\Omega\).

In the presence of interlayer coupling, (normal state) dispersion changes to \(\epsilon_{b\vect k}\). The number of bands stays constant and if the interactions, temperature and chemical potential are small compared to the energy of the lowest energy band (in absolute value) of \(B\backslash\Omega\) in the non-interacting case, the index set \(B\) can be chosen so that the bands in \(B\backslash\Omega\) that were full (empty) in the non-interacting case, are still full (empty) in the interacting case. The interacting charge density is
\begin{equation}
\tilde n(\mu) = \frac {2e}{V} \sum_{b\in B,\vect k} f(\epsilon_{b\vect k}-\mu) = \frac{2e} V \sum_{b\in\Omega,\vect k} f(\epsilon_{b\vect k}-\mu) + n_{\rm high},
\end{equation}
where \(f\) is the Fermi-Dirac distribution at temperature \(T\) and \(n_{\rm high}\) has the the same value as in Eq. \eqref{eq:non-int_charge}. The above has been formulated in the non-linearized theory. To calculate the excess charge relative to the charge neutrality point in the linearized theory, we split the bands between the two valleys and find 
\begin{align}
n(\mu) &\coloneqq \tilde n(\mu) - n_0 = \frac {2e}{V} \sum_{\rho,b\in\Omega,\vect k} \left[ f(\epsilon_{\rho b \vect k}-\mu) - f_0(\epsilon_{0,\rho b \vect k}) \right] \notag\\
    &= 2e\sum_{\rho,b\in\Omega} \int \frac{\dd{\vect k}}{(2\pi)^2} \left[ f(\epsilon_{\rho b \vect k}-\mu) - f_0(\epsilon_{0,\rho b \vect k}) \right],
\end{align}
where \(n\) is the excess charge density and \(\Omega\) is now the set of bands in one valley.

The charge neutrality point \(\mu^*\) is determined from the equation \(n(\mu^*) = 0\). It is shown for different twist angles in Figs. 2(g--i) of the main text, and is always located in the middle between the two DOS peaks.

\section{Simplified model of the superconducting state}
The notion of weak or absent electron--phonon mediated superconductivity in pristine graphene is widely known. Here we reconcile this notion with our results claiming that a quite simple BCS-style model could describe the observations of superconductivity in twisted bilayer graphene. These results are not new, but we follow especially the treatments in Refs. \onlinecite{Kopnin2008,kopninbook2014} and adopt to the notation of the main paper, along with some estimates. 

We start from the generic self-consistency equation for the mean-field order parameter $\Delta$. If $\Delta$ is position independent, the Bogoliubov--de Gennes equation can be solved to yield
\begin{equation}
\Delta = 4 \lambda \int^{\kc}\!\frac{\dd{\vect{k}}}{(2\pi)^2} \frac{\Delta}{E_{\vect{k}}} \tanh\left(\frac{E_{\vect{k}}}{2 \kB T}\right),
\label{eq:simplifiedsc}
\end{equation}
where the prefactor 4 comes from summation over the valley and band indices, where in the band sum we include only the doubly degenerate lowest positive energy band. The cutoff $\kc$ is specified more below. We moreover assume that $E_{\vect{k}}=\sqrt{\epsilon_{\vect{k}}^2+\Delta^2}$. Here and below, without loss of generality we assume $\Delta=|\Delta| \ge 0$. Our idea is to solve the self-consistency equation in three cases: (i) at the Dirac point for a Dirac dispersion $\epsilon_{\vect{k}}^2=\hbar^2 \vF^2 k^2$, (ii) for a Dirac dispersion at non-zero doping $\mu$, i.e., $\epsilon_{\vect{k}}^2 = (\hbar \vF k-\mu)^2$, and (iii) for a flat band with and without doping, $\epsilon_{\vect{k}} \approx \mu$ for $\vect{k}\in \Omega_\text{FB}$. In each case we have the normal-state solution $\Delta=0$, which we exclude by dividing both sides in Eq.~\eqref{eq:simplifiedsc} by $\Delta$. 

Note that Eq.~\eqref{eq:simplifiedsc} {\em does not} represent the full self-consistency equation solved in the main text. Rather, we use it here simply to provide estimates of the behavior of $\Delta$ in various limits.

\subsection{Linear dispersion, no doping}
Far away from the magic angle, the twisted bilayer behaves as if the two graphene layers would be almost uncoupled. This means that the low-energy dispersion exhibits two separate copies of the graphene Dirac dispersion. Inserting an ultraviolet energy cutoff $\epsilonc = \hbar \vF \kc$ and performing the integral for the $T=0$ gap function, the self-consistency equation goes to the form
\begin{equation}
\frac{\pi\hbar^2\vF^2}{2\lambda} = -\Delta + \sqrt{\Delta^2+\epsilonc^2}
\end{equation}
or
\begin{equation}
\Delta = \frac{\pi \hbar^2 \vF^2}{4} \frac{\lambda^2-\lambdac^2}{\lambda \lambdac^2},
\end{equation}
where $\lambdac = \pi\hbar^2\vF^2/(2\epsilonc)$. Since $\Delta \ge 0$, this solution makes sense only if $\lambda > \lambdac$, and otherwise the only possible solution is the normal state $\Delta=0$. 

In pristine graphene, the critical interaction strength can be written also in terms of the nearest-neighbour hopping term $\gamma_0 \approx \SI{3}{\eV}$ \cite{geimreview2009}. Namely, within a nearest-neighbour tight-binding model the Fermi speed of graphene is $\vF = \sqrt{3} \gamma_0 a/(2 \hbar)$, where $a$ is the graphene lattice constant. We hence get
\begin{equation}
\lambdac = \frac{3 \pi}{8} \frac{\gamma_0}{\epsilonc} \gamma_0 a^2.
\end{equation}
If the attractive interaction results from electron--phonon coupling, a typical cutoff energy could be of the order of the Debye energy $\SI{200}{\meV}$ \cite{efetov2010}. In this case $\lambdac \approx \SI{50}{\eV}a^2$, 5 to 50 times larger than the values of $\lambda$ used in our work. 100 to 200 meV is also the range of the maximum cutoff energy that we have used in our numerical results when including the contribution from higher bands. Even if the cutoff $\epsilonc$ would be of the order of $\gamma_0$, the resulting $\lambdac$ would be one order of magnitude larger than the smallest $\lambda$ used in our results.

\subsection{Linear dispersion, with doping}
Let us try to reconcile the observations of superconductivity in Li or Ca doped graphene with the above idea. These cases are more accurately described by \cite{profeta2012} within the Eliashberg theory. Here we just show in which sense doping fits into the above picture. Assuming $\epsilon_{\vect{k}}=  \pm \hbar v_F k-\mu$ and $\Delta < \epsilonc$, and cutting the integral at $\epsilon_{\vect{k}}=\epsilonc$ the self-consistency equation at $T=0$ becomes \cite{Kopnin2008}
\begin{equation}
\frac{\pi\hbar^2\vF^2}{2\lambda} = \sqrt{\Delta^2 + \epsilonc^2}-\sqrt{\Delta^2+\mu^2} + |\mu| \ln \frac{|\mu|+\sqrt{\Delta^2+\mu^2}}{\Delta}.
\end{equation}
Let us assume that $\Delta \ll |\mu|,\epsilonc$ so that we can expand the right hand side in $\Delta$. In this case we find an analytic solution for $\Delta$,
\begin{equation}
\Delta=2|\mu| \exp\left[-\frac{\epsilonc}{|\mu|}\left(\frac{\lambdac}{\lambda}-1\right)-1\right].
\end{equation}
Let us assume a cutoff energy $\epsilonc=\SI{200}{\meV}$ and a coupling strength $\lambda=\lambdac/22$ (corresponding to about $\SI{5}{\eV}a^2$ with the above estimates). In this case, with $\mu=\SI{0.7}{\eV}$ we would get $\Delta=\SI{1.3}{\meV}$. This corresponds to a critical temperature of $\SI{9}{K}$, in the same range as the one that was measured in Li or Ca doped graphene \cite{ludbrook2015evidence, tiwari2017superconductivity, Chapman2016, ichinokura2016superconducting}.

\subsection{Flat band estimate}
Let us now make similar estimates for the flat-band case of the \moire superlattice. In this case, we assume that $\Delta$ is \emph{larger} than the bandwidth of the lowest-energy band. Within that band, we can hence approximate $E_{\vect{k}} \approx \Delta$ in Eq.~\eqref{eq:simplifiedsc} and at $T=0$ the integral is over a constant function. As a result, we get
\begin{equation}
\Delta_\text{FB} = \frac{\lambda}{\pi^2} \Omega_\text{FB} = \frac{8\lambda}{\sqrt{3}(3m^2+3m+1)a^2},
\end{equation}
where $\Omega_\text{FB} = 8\pi^2/[\sqrt{3}(3m^2+3m+1)a^2]$ is the area of the first Brillouin zone of the \moire superlattice. Within the model adapted in the main text, the magic angle is around $m \approx 34$, in which case we would get $\Delta_\text{FB} = 1.3 \times 10^{-3} \lambda/a^2$. In Fig.~3b of the main text, the solid line has a slope of $1.0 \times 10^{-3} \lambda/a^2$, \ie, very close to this simple estimate.

The temperature dependent $\Delta$ in the flat-band case is obtained by solving
\begin{equation}
\Delta=\Delta_\text{FB} \tanh\left(\frac{\Delta}{2 \kB T}\right).
\end{equation}
At the critical temperature, $\Delta \rightarrow 0$, and we can hence expand the right hand side to the linear order in $\Delta/(2\kB\Tc)$. This directly yields $\kB\Tc=\Delta_\text{FB}/2$. 

In the case of a non-zero potential $\mu$, we can use $E_{\vect{k}} \approx \sqrt{\mu^2+\Delta^2}$ in the self-consistency equation. It then becomes (for $\Delta >0$)
\begin{equation}
\Delta=\Delta_\text{FB} \frac{\Delta}{\sqrt{\mu^2 + \Delta^2}}, \text{ or } \Delta=\sqrt{\Delta_{\rm FB}^2-\mu^2}.
\end{equation}
In this case superconductivity is hence suppressed when the absolute value of the chemical potential is larger than $\Delta_{\rm FB}$.

\section{Simplified model for Coulomb pseudopotential}

Coulomb interaction differs from the electron--phonon interaction due to the fact that photons are almost instantaneous, whereas for phonon-mediated interaction we have to take the retardation into account. Usually in BCS theory, and also in our model, we approximate the retardation by imposing an energy cutoff at the maximum phonon frequency \(\omega_D\) in the self-consistency equation. For Coulomb interaction there is no physical cutoff, and consequently, we cannot operate in purely low-energy regime. The high energy states do contribute logarithmically to \(\Delta\) at low energies.

The proper way to formulate the low-energy theory with a cutoff which also applies to the Coulomb interaction, is to define a modified pseudopotential \(u^*\) which replaces the bare interaction in the self-consistency equation and takes the high-energy parts into account. If \(\Delta(\vect r)\) is position-dependent, the pseudopotential will be a matrix of two position coordinates \(u^*(\vect r, \vect r')\). If \(\Delta\) is constant in space, the pseudopotential is a scalar.

We want to consider the effect of the Hubbard interaction, described in the continuum limit by the Hamiltonian
\begin{equation}
H_\text{Hubbard} = \frac{u}{2} \sum_{\mathclap{\sigma,\alpha,i\,}} \int\dr \psi_{\sigma,\alpha i}^\dagger(\vect{r}) \psi_{\bar\sigma,\alpha i}^\dagger(\vect{r})	\psi_{\bar\sigma,\alpha i}(\vect{r}) \psi_{\sigma,\alpha i}(\vect{r}),
\end{equation}
where \(u=Ua^2\) and \(U\) is the Hubbard parameter describing the on-site interaction in the tight-binding model. We assume that \(U>0\) so that the interaction is repulsive. The inclusion of such an interaction has multiple effects in a inhomogeneous system, but here we only consider the effect on the order parameter through the modification of the self-consistency equation.

As we are now not doing a low-energy calculation, separation into valleys is not useful and we cannot do the continuum approximation in which we assume that the graphene lattice \(L\) is duplicated infinitely many times in the superlattice. Therefore, in the following the sums and integrals are done over the sets \(\vect G\in SL^*/L^*\) and \(\vect k\in\R^2/SL^*\). The two graphene valleys are then separated from each other by a large, but finite \(\vect G\)-vector. The valley sum is thus included in the sum over \(\vect G\) and there is no valley index \(\rho\). %As a sanity check, the combined sum-integral over \(\vect G\) and \(\vect k\) spans the area of the reciprocal graphene lattice.

For simplicity, we assume \(\Delta_{\alpha i}(\vect G) = \Delta \delta_{\vect G,0}\) so that \(\Delta\) has no position dependence and is the same on both layers and sublattices. With this simplification, we can diagonalize the Hamiltonian \(\underline{\mathcal{H}}_{\vect k}\) and the order parameter simultaneously in the BdG equation [Eq.~\eqref{eq:bdg_kspace} of the main text], which we write as
\begin{align}
    \begin{pmatrix}
		\underline{\mathcal{H}}_{\vect k} & {\Delta}\underline{1} \\
        {\Delta}^*\underline{1} & -\underline{\mathcal{H}}_{\vect k}
	\end{pmatrix}
    \begin{pmatrix}
    		\underline{u}_{\vect{k}b} \\ \underline{v}_{\vect{k}b}
    \end{pmatrix}
    =
	E_{\vect{k}b}
	\begin{pmatrix}
		\underline{u}_{\vect{k}b} \\ \underline{v}_{\vect{k}b}
	\end{pmatrix},
\end{align}
where the underlined quantities are matrices/vectors with indices \(\vect G, \alpha\), and \(i\). Let \(\underline{\mathcal{G}}_{\vect k}\) be a unitary transformation which diagonalizes the normal state Hamiltonian \(\underline{\mathcal{H}}_{\vect k}\). Then the above equation becomes
\begin{align}
    \begin{pmatrix}
		\underline{\epsilon}_{\vect k} & {\Delta}\underline{1} \\
        {\Delta}^*\underline{1} & -\underline{\epsilon}_{\vect k}
	\end{pmatrix}
    \begin{pmatrix}
    	\underline{u}_{\vect{k}b}' \\
        \underline{v}_{\vect{k}b}'
    \end{pmatrix}
    =
	E_{\vect{k}b}
	\begin{pmatrix}
    	\underline{u}_{\vect{k}b}' \\
        \underline{v}_{\vect{k}b}'
	\end{pmatrix},
\end{align}
where \(\underline{u}_{\vect{k}b}' = \underline{\mathcal{G}}_{\vect k}\underline{u}_{\vect{k}b}\), \(\underline{v}_{\vect{k}b}' = \underline{\mathcal{G}}_{\vect k}\underline{v}_{\vect{k}b}\) and \(\underline{\epsilon}_{\vect k}=\underline{\mathcal{G}}_{\vect k} \underline{\mathcal{H}}_{\vect k}\underline{\mathcal{G}}_{\vect k}^\dag\). We now label the normal state eigenstates with band index \(b\). With constant \(\Delta\), the positive-energy BdG eigenstates are in simple correspondence with the eigenstates (both positive and negative energy) of the normal state, and can also be labeled with the same indices. Concentrating to a single Nambu-block of the BdG equation, 
\begin{align}
    \begin{pmatrix}
		{\epsilon}_{\vect kb} & {\Delta} \\
        {\Delta}^* & -{\epsilon}_{\vect kb}
	\end{pmatrix}
    \begin{pmatrix}
    		{u}_{\vect{k}b}' \\ {v}_{\vect{k}b}'
    \end{pmatrix}
    =
	E_{\vect{k}b}
	\begin{pmatrix}
		{u}_{\vect{k}b}' \\ {v}_{\vect{k}b}'
	\end{pmatrix},
\label{eq:bdg_matrix_diagonal}
\end{align}
we find that the eigenenergies and eigenstates assume the usual BCS form 
\begin{align}
E_{\vect k b} &= \sqrt{\epsilon_{\vect k b}^2 + \abs{\Delta}^2},\\
u_{\vect k b} &= \frac{e^{i\phi}}{\sqrt 2} \left( 1 + \frac{\epsilon_{\vect kb}}{E_{\vect k b}} \right)^{1/2},\\
v_{\vect k b} &= \frac{1}{\sqrt 2} \left( 1 - \frac{\epsilon_{\vect kb}}{E_{\vect k b}} \right)^{1/2},
\end{align}
where \(\phi=\arg(\Delta)\).

The self-consistency equation [Eq.~\eqref{eq:self-consistency_equation_kspace} in the main text with \(\vect G=0\) and generalized to include energy-dependent interactions] can be written in the above matrix notation as
\begin{align}
	\Delta_{\vect kb,\alpha i} = & \sum_{b'} \int\!\frac{\dd{\vect{k}'}}{(2\pi)^2} V_{\vect k\vect k'}^{b b'} \left(\underline{u}_{\vect{k}' b'}^\dag  \underline{\Pi}_{\alpha i} \underline{v}_{\vect{k}'b'}\right)^* \tanh(\frac{E_{\vect{k}'b'}}{2\kB T}),
\end{align}
where \(\underline{\Pi}_{\alpha i}\) is the projection operator to the sublattice \(\alpha\) and layer \(i\). We assume that the interaction has the simplified BCS form
\begin{equation}
V_{\vect k\vect k'}^{bb'} = \lambda \theta(\abs{\epsilon_{\vect k b}} - \omega_D)\theta(\abs{\epsilon_{\vect k' b'}} - \omega_D) - u,
\end{equation}
with electron--phonon cutoff at Debye energy \(\omega_D\).

The sum of complete set of projection operators is an identity: \(\sum_{\alpha, i}\underline{\Pi}_{\alpha i} = \underline 1\). To get rid of the projection operator, we take the average over \(\alpha\) and \(i\). As \(\Delta_{\vect kb,\alpha i} = \Delta_{\vect kb}\), we get
\begin{align}
	\Delta_{\vect kb} &= \frac{1}{4} \sum_{b'} \int\!\frac{\dd{\vect{k'}}}{(2\pi)^2} V_{\vect k\vect k'}^{bb'}\left(\underline{u}_{\vect{k}' b'}^\dag \underline{v}_{\vect{k}'b'}\right)^* \tanh(\frac{E_{\vect{k}'b'}}{2\kB T}) \notag\\
			&=\frac{1}{4} \sum_{b'} \int\!\frac{\dd{\vect{k}'}}{(2\pi)^2}  V_{\vect k\vect k'}^{bb'} {u}_{\vect{k}' b'}' ({v}_{\vect{k}'b'}')^* \tanh(\frac{E_{\vect{k}' b'}}{2\kB T}).
			\label{eq:sc_simple}
\end{align}
In the second line, we did a basis transformation with the matrix \(\underline{\mathcal{G}}_{\vect p}^\dag\).

We now divide \(\Delta_{\vect{k} b} = \Delta_{\vect{k} b}^\lambda + \Delta^u\) into two parts, with \(\Delta_{\vect k b}^\lambda\) corresponding to the \(\lambda\) part of the interaction in the RHS of Eq.~\eqref{eq:sc_simple} and \(\Delta^\text{u}\) corresponding to the \(u\) part of the interaction \cite{morel1962calculation}. The difference between the two terms is in the energy dependence. \(\Delta_{\vect{k} b}^\lambda\) vanishes above the cutoff, but \(\Delta^u\) has no energy dependence and persists at high energies. With this division, the self-consistency equation splits into two coupled equations,

\begin{align}
	\Delta_{\vect kb}^\lambda &= \phantom{+} \frac{\lambda}{4} \sum_{b'} \int_{\abs{\epsilon_{\vect k'b'}} < \omega_D}\!\frac{\dd{\vect{k}'}}{(2\pi)^2} {u}_{\vect{k}' b'}' ({v}_{\vect{k}'b'}')^* \tanh(\frac{E_{\vect{k}' b'}}{2\kB T}) \notag\\
	&\qquad\qquad\qquad\qquad\times \theta(\abs{\epsilon_{\vect k b}} - \omega_D),\label{eq:delta_lambda} \\
	\Delta^u &= - \frac{u}{4} \sum_{b'} \int_{\abs{\epsilon_{\vect k'b'}} < \omega_D}\!\frac{\dd{\vect{k}'}}{(2\pi)^2} {u}_{\vect{k}' b'}' ({v}_{\vect{k}'b'}')^* \tanh(\frac{E_{\vect{k}' b'}}{2\kB T}) \notag\\
	&\phantom{=} - \frac{u}{4} \sum_{b'} \int_{\abs{\epsilon_{\vect k'b'}} > \omega_D}\!\frac{\dd{\vect{k}'}}{(2\pi)^2} {u}_{\vect{k}' b'}' ({v}_{\vect{k}'b'}')^* \tanh(\frac{E_{\vect{k}' b'}}{2\kB T}).
	\label{eq:delta_u}
\end{align}
Above, we also split the sums and integrals over the eigenstates to low and high energy parts with \(\omega_D\) as the cutoff. Assuming \(\omega_D \gg T, \Delta^u\), we can approximate that for high energy states 
\begin{align}
u_{\vect kb}' (v_{\vect kb}')^* \tanh(\frac{E_{\vect{k} b}}{2\kB T}) \approx \frac{\Delta^u}{2\abs{\epsilon_{\vect kb}}}.
\end{align}
Inserting this into Eq.~\eqref{eq:delta_u}, we can (partially) solve for \(\Delta^u\) to obtain an equation which only refers to the low energy states,
\begin{equation}
\Delta^u = - \frac{u^*}{4} \sum_{b'} \int_{\abs{\epsilon_{\vect k'b'}} < \omega_D}\!\frac{\dd{\vect{k}'}}{(2\pi)^2} {u}_{\vect{k}' b'}' ({v}_{\vect{k}'b'}')^* \tanh(\frac{E_{\vect{k}' b'}}{2\kB T}).
\end{equation}
The high energy states renormalize the interaction constant, which is replaced by the Coulomb pseudopotential
\begin{equation}
u^*=\frac{u}{1 + u \alpha},
\end{equation}
where
\begin{equation}
\alpha = \frac{1}{4} \sum_{b} \int_{\abs{\epsilon_{\vect k'b'}} > \omega_D}\!\frac{\dd{\vect{k}}}{(2\pi)^2} \frac{1}{2\abs{\epsilon_{\vect kb}}}.
\end{equation}
The equation for the full order parameter, including both interactions, is now
\begin{equation}
	\Delta =  \frac{\lambda_\text{eff}}{4} \sum_{b'} \int_{\abs{\epsilon_{\vect k'b'}} < \omega_D}\!\frac{\dd{\vect{k}'}}{(2\pi)^2} {u}_{\vect{k}' b'}' ({v}_{\vect{k}'b'}')^* \tanh(\frac{E_{\vect{k}' b'}}{2\kB T})
	\label{eq:delta_final}
\end{equation}
with
\begin{equation}	
  \lambda_\text{eff} = \lambda-u^*.
\end{equation}
If \(\Delta(\vect r)\) is position dependent, the derivation becomes more complicated, and in the end, the pseudopotential becomes a matrix \(u^*(\vect G, \vect G')\) instead of a scalar like above.

\begin{figure}
    \includegraphics[width=0.9\columnwidth]{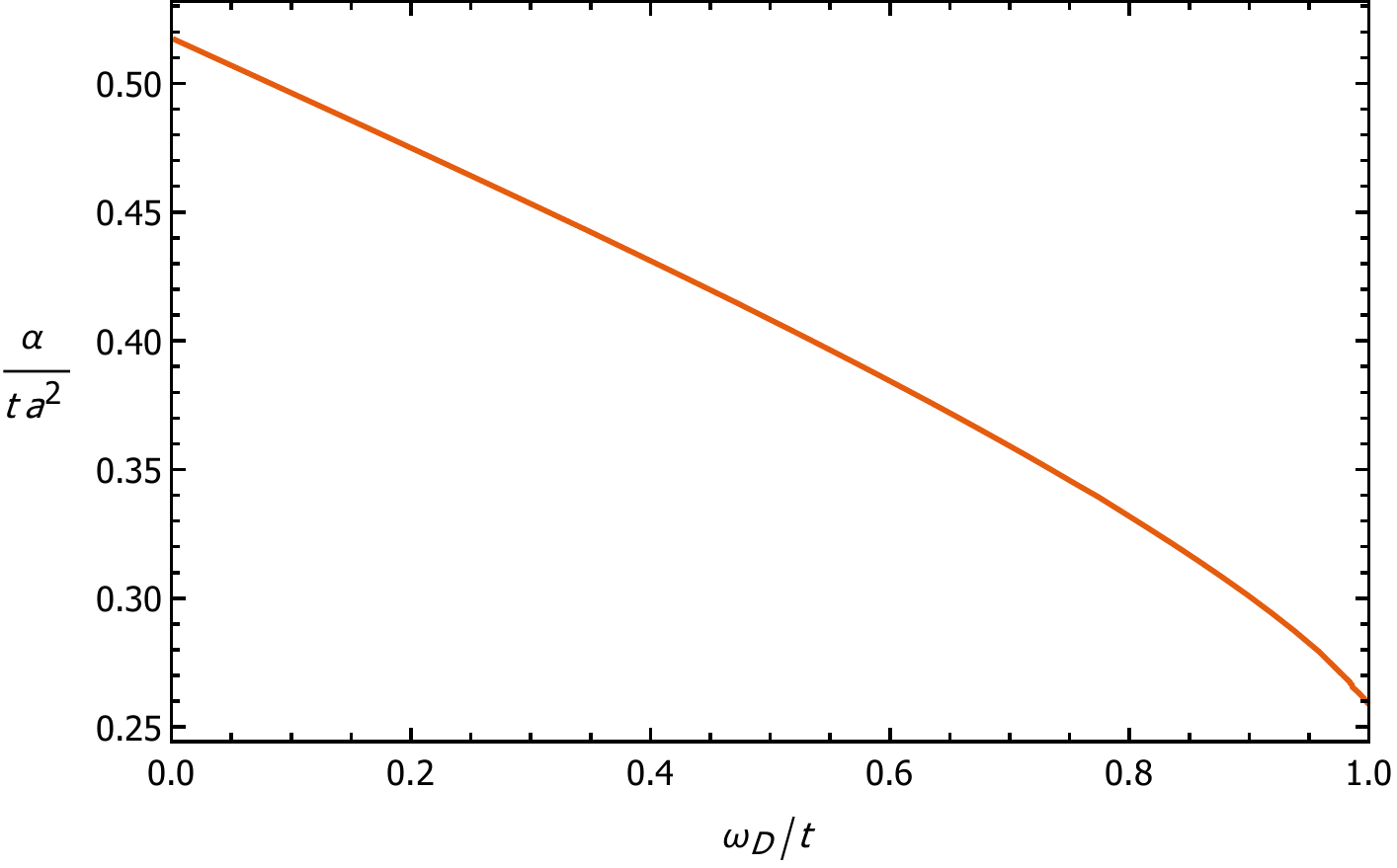}
    \caption{Dependence of the graphene pseudopotential renormalization constant \(\alpha\) on the electron--phonon cutoff \(\omega_D\). Pseudopotential renormalization constant of TBG can be approximated with that of graphene if \(\abs{t_\perp}\ll\omega_D\).}
    \label{fig:pseudopotential}
\end{figure}

The pseudopotential renormalization parameter \(\alpha\) now depends on structure of the high energy bands. It is not very sensitive to the parameters of the system and for this calculation we assume \(t_\perp = 0\) so that the two graphene layers are completely independent of each other. The sums and integrals then transform as
\begin{align}
  \alpha &= \frac{1}{4}\sum_{b} \int_{\substack{\R^2/SL^*\\\abs{\epsilon_{\vect kb}} > \omega_D}}\!\frac{\dd{\vect{k}}}{(2\pi)^2} \frac{1}{2\abs{\epsilon_{\vect kb}}} \notag\\
  &\approx \frac{1}{4}\sum_i \sum_{b\in\pm1} \int_{\substack{\R^2/L^*\\\abs{\epsilon_{\vect kb}} > \omega_D}}\!\frac{\dd{\vect{k}}}{(2\pi)^2} \frac{1}{2\abs{\epsilon^0_{\vect kb}}},
\end{align}
where \(\epsilon^0_{\vect k b}\) are the graphene eigenenergies calculated from the tight binding model with only nearest neighbour hoppings.

If approximated as above, \(\alpha\) corresponds to the pseudopotential constant for graphene. We show the dependence on the cutoff \(\omega_D\) in Fig.~\ref{fig:pseudopotential}. With parameters \(\omega_D=\SI{200}{meV}\) and nearest neighbour hopping \(t=\SI{3}{eV}\), we find that \(\alpha \approx \SI{0.2}{eV^{-1}}a^{-2}\). The maximum value for the pseudopotential is thus \(u^*_\text{max} = 1/\alpha \approx \SI{5}{eV} a^2\), which is obtained in the limit \(U\to\infty\). For \(U=\SI{5}{eV}\), the effective interaction strength is reduced to half of the bare interaction strength, \(u^* \approx 0.5 u = \SI{2.5}{eV}a^2\).

\end{document}